\documentclass[aps,prb,twocolumn,groupedaddress,showpacs]{revtex4}

\usepackage{bm}
\usepackage{graphicx}

\usepackage{color}

\begin{document}

\title{Angle-resolved photoemission spectroscopy of Co-based boride superconductor LaCo$_{1.73}$Fe$_{0.27}$B$_2$}

\author{
	K. Nakayama,$^{1}$
	E. Ieki,$^{1}$
	Y. Tanaka,$^{1}$
	T. Sato,$^{1,2}$
	T. Takahashi,$^{1,3}$
	T. Kuroda,$^{4}$
	H. Mizoguchi,$^{5}$
	S. W. Kim,$^{5,\ast}$
	and H. Hosono$^{4,5}$}

\affiliation{$^1$Department of Physics, Tohoku University, Sendai 980-8578, Japan}
\affiliation{$^2$TRiP, Japan Science and Technology Agency (JST), Kawaguchi 332-0012, Japan}
\affiliation{$^3$WPI Research Center, Advanced Institute for Materials Research, Tohoku University, Sendai 980-8577, Japan}
\affiliation{$^4$Materials and Structures Laboratory, Tokyo Institute of Technology, Yokohama 226-8503, Japan}
\affiliation{$^5$Frontier Research Center, Tokyo Institute of Technology, Yokohama 226-8503, Japan}

\date{\today}

\begin{abstract}
We have performed angle-resolved photoemission spectroscopy of Co-based boride superconductor LaCo$_{1.73}$Fe$_{0.27}$B$_2$ ($T_c$ = 4.1 K), which is isostructural to the 122-type Fe-pnictide superconductor with the pnictogen atom being replaced with boron.  We found that the Fermi level is located at a dip in the density of states (DOS) in contrast to Co-pnictide ferromagnets.  This reduction in DOS together with the strong Co 3$d$-B 2$p$ covalent bonding removes the ferromagnetic order and may cause the superconductivity.  The energy bands near the Fermi level show higher three dimensionality and a weaker electron-correlation effect than those of Fe pnictides.  The Fermi surface topology is considerably different from that of Fe pnictides, suggesting the difference in the superconducting mechanism between boride and pnictide superconductors.
\end{abstract}

\pacs{74.70.Xa, 74.25.Jb, 79.60.-i}

\maketitle

\section{Introduction}
Realization of superconductivity in materials containing ferromagnetic elements such as Fe, Co, and Ni has been one of the most challenging issues in solid-state physics, because the magnetic element has been regarded to destroy the superconductivity.  The discovery of high-temperature superconductivity in F-doped LaFeAsO \cite{Kamihara} is quite surprising in this regard and opens a new avenue to the superconductivity research.  LaFeAsO is composed of alternately stacked LaO and FeAs layers, and the structural engineering with these blocks has brought about a variety of Fe-pnictide and Fe-chalcogenide superconductors such as BaFe$_2$As$_2$, LiFeAs, Sr$_4$V$_2$O$_6$Fe$_2$As$_2$, FeSe, and K$_x$Fe$_{2-y}$Se$_2$ \cite{FeReview}.  Further search for novel superconductors has been performed by replacing Fe atoms with other transition metals, leading to the discovery of Ni-pnictide superconductors \cite{LaNiPO, LaNiAsO, BaNi2P2, BaNi2As2}.  However, despite intensive efforts, Co-based superconductors have been hardly discovered unlike the case of Fe or Ni, probably due to the relatively stronger ferromagnetic correlation.  In fact, LaCoAsO, LaCoPO, LaCo$_2$As$_2$, and LaCo$_2$P$_2$ undergo the ferromagnetic transition with no signature of superconductivity \cite{LaCoAsO, LaCo2As2, LaCo2P2}.

Recent success in synthesizing Co-based superconductor La(Co$_{1-x}$Fe$_x$)$_2$B$_2$ has provoked much attention, because LaCo$_2$B$_2$ is isostructural to BaFe$_2$As$_2$ \cite{Mizoguchi} with As atoms being replaced with boron (B).  Remarkable aspects of LaCo$_2$B$_2$ are (i) the FeAs layer in the pnictide superconductor is completely replaced with the CoB layer, (ii) the non-doped parent LaCo$_2$B$_2$ shows Pauli paramagnetism even though it has the same crystal structure as LaCo$_2$As$_2$ and LaCo$_2$P$_2$ ferromagnets, and (iii) the superconductivity with a transition temperature ($T_c$) of $\sim$4 K emerges upon hole doping into the parent compound, while the electron-doped counterpart is not superconducting \cite{Mizoguchi}.  To elucidate the microscopic mechanism of the peculiar magnetic and superconducting properties as well as the role of the B replacement, it is of particular importance to clarify the electronic structure near the Fermi level ($E_{\rm F}$).  Although a previous theoretical calculation suggested a dominant contribution from the Co 3$d$ orbital to the electronic states at $E_{\rm F}$ \cite{Mizoguchi}, a different calculation pointed out an additional significant contribution from the La 5$d$ states \cite{Wang}.  Such a controversy strongly requests experimental clarification of the electronic structure of this novel material.

In this article, we report the first direct observation of the electronic structure of LaCo$_{1.73}$Fe$_{0.27}$B$_2$ superconductor by angle-resolved photoemission spectroscopy (ARPES) with synchrotron radiation.  Using the tunable photon energy, we have determined the electronic structure in the three-dimensional momentum space and compared it with the band calculations.  We have revealed a drastic chemical-potential shift in LaCo$_{1.73}$Fe$_{0.27}$B$_2$ with respect to that of ferromagnetic Co pnictides, the highly asymmetric density of states (DOS) relative to $E_{\rm F}$, and the dominant Co 3$d$ nature of the electronic states near $E_{\rm F}$.  We also found the coexistence of two- and three-dimensional Fermi-surface (FS) sheets, together with the moderate mass-renormalization of bands near $E_{\rm F}$.  We discuss implications of observed electronic structure to the origin of characteristic physical properties of this Co-based boride superconductor.

\section{EXPERIMENTS}
High-quality single crystals of LaCo$_{1.73}$Fe$_{0.27}$B$_2$ ($T_c$ = 4.1 K) were grown by the floating zone method, and the Fe content was determined by electron-probe microanalysis (EPMA) measurements.  The crystal structure of La(Co$_{1-x}$Fe$_x$)$_2$B$_2$ is shown in the inset of Fig. 1.  It consists of La and (Co$_{1-x}$Fe$_x$)B layers stacked alternately along the $c$ axis, forming the tetragonal ThCr$_2$Si$_2$ structure as in BaFe$_2$As$_2$.  High-resolution ARPES measurements were performed at BL-28A in Photon Factory, KEK, Tsukuba, using a VG-SCIENTA SES2002 spectrometer.  All the APRES data have been measured at 30 K with circularly polarized photons of 36-130 eV.  The energy and angular resolutions were set at 15-30 meV and 0.2$^{\circ}$, respectively.  Clean surfaces for ARPES measurements were obtained by cleaving crystals $in$ $situ$ in a vacuum better than 1$\times$10$^{-10}$ Torr.

\section{RESULTS AND DISCUSSION}
\begin{figure}[b]
\begin{center}
\includegraphics[width=2.5in]{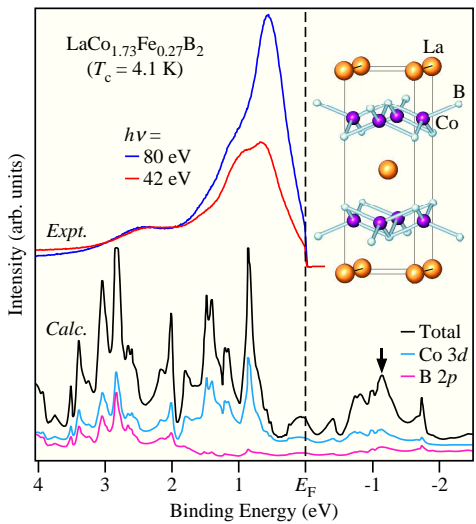}
\end{center}
\caption{
(color online) Valence-band DOS of LaCo$_{1.73}$Fe$_{0.27}$B$_2$ ($T_c$ = 4.1 K) obtained by integrating the ARPES spectrum over the momentum region enclosed by the $\Gamma$(Z)-M(A)-X(R) triangle [see Fig. 2(a)] at 30 K with $h\nu$ = 42 (red curve) and 80 eV (blue curve).  The calculated partial DOS of Co 3$d$ and B 2$p$ orbitals (light blue and pink curves, respectively) and the total DOS (black curve) in the parent LaCo$_2$B$_2$ (ref. 10) are also plotted in the bottom.  The inset shows the crystal structure of LaCo$_2$B$_2$.
}
\end{figure}

\begin{figure*}[t]
\begin{center}
\includegraphics[width=5in]{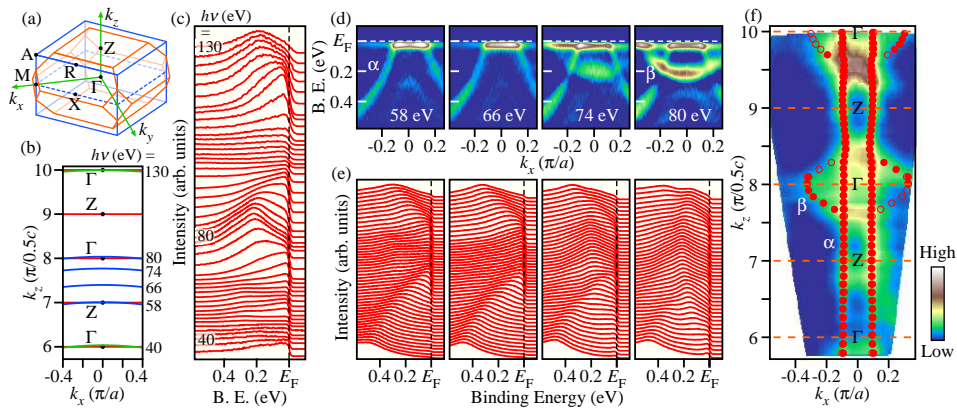}
\end{center}
\caption{
(color online) (a) Three-dimensional Brillouin zone of LaCo$_{1.73}$Fe$_{0.27}$B$_2$ (orange line) and its simplified version (blue line).  (b) Brillouin zone in the $k_x$-$k_z$ plane.  (c) Normal-emission ARPES spectra near $E_{\rm F}$ measured at 30 K with various photon energies.  (d) Photon energy dependence of second-derivative ARPES intensity and (e) corresponding ARPES spectra.  Their momentum cuts in the $k_x$-$k_z$ plane are indicated by blue curves in (b).  (f) ARPES intensity at $E_{\rm F}$ plotted as a function of $k_x$ and $k_z$.  The intensity was obtained by integrating the ARPES intensity at $E_{\rm F}$$\pm$10 meV.  Filled and open circles represent the Fermi wave vector ($k_{\rm F}$) points determined by tracing the experimental band dispersions and the crystal symmetry.
}
\end{figure*}

Figure 1 shows the valence-band DOS of LaCo$_{1.73}$Fe$_{0.27}$B$_2$ obtained by integrating the ARPES spectrum at 30 K over the whole Brillouin zone.  There are several characteristics such as a monotonic decrease of the DOS toward $E_{\rm F}$, a prominent peak at $\sim$0.5 eV with a shoulder at $\sim$1.0 eV, and a small hump at $\sim$2.5 eV.  Comparison with the theoretical DOS calculated for the parent LaCo$_2$B$_2$ \cite{Mizoguchi} reveals that the electronic states at $E_{\rm F}$-1 eV and at around 2.5 eV are attributed to the Co 3$d$ dominated states and the B 2$p$ and Co 3$d$ hybridized states, respectively.  This assignment is consistent with the experimental fact that the spectral intensity at $E_{\rm F}$-1 eV is relatively enhanced with respect to that at 2.5 eV in the measurement with 80-eV photons, because the photoionization cross-section ratio Co 3$d$/B 2$p$ is about five times larger in $h\nu$ = 80 eV than in 42 eV \cite{Lindau}.  We also notice a finite difference in the peak position between the experimental and calculated DOS, which is probably due to the chemical potential shift caused by the hole doping with the partial Fe substitution and the possible mass-renormalization originating from the electron correlation.  The present results solve the controversy regarding the electronic states of LaCo$_2$B$_2$.  While one of the band calculations has predicted the very high DOS due to the La 5$d$ states at $E_{\rm F}$ \cite{Wang}, the present experimental result does not support this prediction, because the cross-section ratio La 5$d$/B 2$p$ is more than two times smaller in $h\nu$ = 80 eV than in 42 eV \cite{Lindau} in sharp contrast to the strong enhancement of spectral intensity near $E_{\rm F}$ at $h\nu$ = 80 eV.  Alternatively, the present experimental result clearly shows that the valence of La atoms is close to +3 and the metallic conduction occurs in the CoB layer in this compound.

To clarify the momentum dependence of the electronic structure near $E_{\rm F}$, we performed ARPES measurements in the normal-emission configuration as a function of $h\nu$, which allow us to probe the out-of-plane band dispersion.  As seen in Fig. 2(c), we found a dispersive band whose energy position exhibits a marked $h\nu$ dependence, demonstrating a strong $k_z$ dispersion.  This band forms an electron pocket with the bottom at $h\nu$ $\sim$40, 80, and 130 eV.  From the observed periodicity, we evaluated the inner potential value ($V_0$) to be 16 eV and the momentum location of the electron-band bottom at the $\Gamma$ point in the three-dimensional Brillouin zone [see Figs. 2(a) and 2(b)].  The bandwidth along the $k_z$ direction is estimated to be $\sim$0.4 eV by fitting the observed $k_z$ dispersion with a simple cosine function.  This value is much larger than that (at most 0.2 eV) reported for 122-type Fe pnictides \cite{Mannella, Yoshida}, suggesting the enhanced three dimensionality of the electronic structure.  Figures 2(d) and 2(e) show the in-plane band dispersion near $k_x$ = 0 measured at several different $k_z$ values.  In addition to the electron band that appears near the $\Gamma$ point, a holelike band intersects $E_{\rm F}$ at $k_x$ $\sim$ 0.1$\pi$/$a$ irrespective of $k_z$ value, suggesting its negligibly small dispersion along the $k_z$ direction.  These experimental results demonstrate the coexistence of a quasi two-dimensional hole pocket ($\alpha$ band) and a strongly three-dimensional electron pocket ($\beta$ band).  The difference in the dimensionality between the two FSs is more visible in the FS mapping in the $k_x$-$k_z$ plane [Fig. 2(f)].

\begin{figure}[b]
\begin{center}
\includegraphics[width=2.5in]{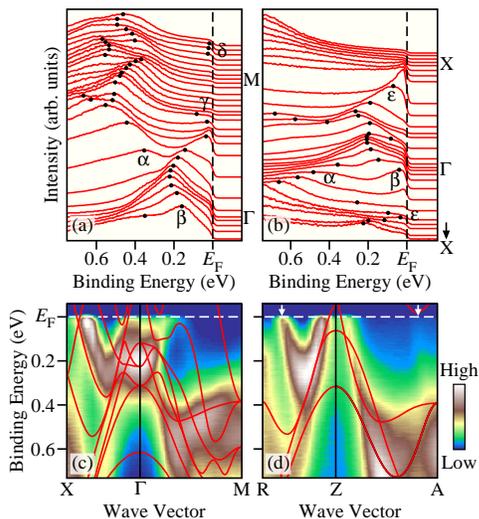}
\end{center}
\caption{
(color online) (a) and (b) Near-$E_{\rm F}$ ARPES spectra measured with $h\nu$ = 80 eV at 30 K along the $\Gamma$-M and $\Gamma$-X high-symmetry lines, respectively.  Dots show the position of peaks in the ARPES spectra.  (c) and (d) ARPES-intensity plot as a function of binding energy and wave vector measured with $h\nu$ = 80 and 62 eV, respectively.  These two photon energies correspond to the X-$\Gamma$-M and the R-Z-A lines, respectively. The calculated band dispersions for the corresponding $k_z$ values are also shown by red curves [$k_z$ = 0 and $\pi$ for (c) and (d), respectively].  The calculated bands for non-doped LaCo$_2$B$_2$ (ref. 10) were renormalized by a factor of 1.5 and shifted upward by 100 meV.
}
\end{figure}

Next we show the in-plane band dispersion at the $\Gamma$MX$\Gamma$ and ZARZ planes (we call $\Gamma$ and Z planes, respectively).  In Figs. 3(a) and 3(b), we show ARPES spectra at 30 K in the $\Gamma$ plane measured with 80-eV photons along the $\Gamma$-M and $\Gamma$-X high-symmetry lines, respectively.  Besides the $\alpha$ and $\beta$ pockets seen in Fig. 2, we notice several characteristic features.  For instance, in the $\Gamma$-M cut, a highly dispersive band ($\gamma$ band), as also visualized in the intensity plot in Fig. 3(c), crosses $E_{\rm F}$ outside the $\beta$ band.  There is a weak but finite Fermi-edge structure near the M point, suggestive of another band crossing ($\delta$ band).  In addition, we find a holelike dispersion ($\epsilon$ band) touching $E_{\rm F}$ near the X point.  As seen in Fig. 3(c), we find a relatively good correspondence between the observed band dispersions and the calculated bands, when the calculated bands \cite{Mizoguchi} are renormalized by a factor of 1.5 so as to obtain the best match of the Fermi velocities of the $\alpha$, $\beta$, $\gamma$, and $\epsilon$ bands.  This renormalization factor is smaller than the value (2-3) reported for Fe-based superconductors \cite{Hong}, indicating the moderate electron correlation in the Co-based compound.  In the Z plane [Fig. 3(d)], the presence (absence) of the $\alpha$ hole ($\beta$ electron) pocket at the zone center agrees well with the calculation.  On the other hand, the experimental band dispersions denoted by white arrows in Fig. 3(d) do not seem to be well reproduced in the calculation.  Since these bands most likely correspond to the $\delta$ and $\epsilon$ bands which cross $E_{\rm F}$ at similar wave vectors in the $\Gamma$ plane [see Figs. 3(a)-3(c)], the present results point to their enhanced two dimensionality as compared to the calculation.

Figures 4(a) and 4(b) display experimental FSs at the $\Gamma$ and Z planes, respectively.  At the zone center, we observe the $\alpha$ and $\beta$ FSs with cylindrical and spherical shapes, respectively, as expected from Fig. 2(f).  Near the X and R points, a quasi-two-dimensional holelike $\epsilon$ FS elongated along the X-M direction is clearly seen.  The $\delta$ band, which is originally a large hole pocket centered at the Z point, forms an electronlike FS with a pinched square shape near the A point due to the hybridization.  The presence of a similar FS near the M point suggests its two-dimensional nature.  In addition, there is a three-dimensional FS arising from the $\gamma$ band whose topology is still unclear at present.  These results unambiguously show that LaCo$_{1.73}$Fe$_{0.27}$B$_2$ is a multiband superconductor with a complicated FS topology due to the presence of multiple FS sheets with different anisotropies and dimensionalities.

\begin{figure}[t]
\begin{center}
\includegraphics[width=2.5in]{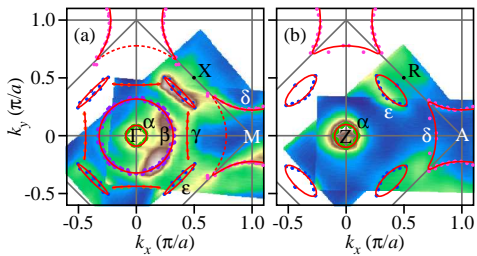}
\end{center}
\caption{
(color online) (a) and (b) ARPES intensity at $E_{\rm F}$ of LaCo$_{1.73}$Fe$_{0.27}$B$_2$ plotted as a function of two-dimensional in-plane wave vector measured with $h\nu$ = 80 and 62 eV, respectively.  The experimentally determined $k_{\rm F}$ points are shown by dots.  Red curves are guides to trace the FSs.
}
\end{figure}

Now we discuss why the ferromagnetism does not appear in LaCo$_{1.73}$Fe$_{0.27}$B$_2$ in contrast to Co pnictides.  We think that this is associated with (i) the chemical potential shift and (ii) the change of the band structure itself.  The point (i) is evident from the comparison of the DOS between ferromagnetic Co pnictides and LaCo$_{1.73}$Fe$_{0.27}$B$_2$.  In the ferromagnetic pnictides such as LaCo$_2$P$_2$, the high DOS due to a prominent peak at $E_{\rm F}$ is expected to stabilize the itinerant ferromagnetic ground state \cite{LaCoAsOcalc, LaCo2P2calc}.  In contrast, this prominent peak structure moves into the unoccupied state in LaCo$_{1.73}$Fe$_{0.27}$B$_2$ because of the hole doping from B atoms (see an arrow in Fig. 1), resulting in the chemical potential being located at a dip of the DOS.  The significant reduction of DOS at $E_{\rm F}$ in LaCo$_{1.73}$Fe$_{0.27}$B$_2$ moves the system away from the ferromagnetic instability.  It is noted that while the chemical potential shift is also caused by partial Fe substitution, the shift expected from the Fe content in our sample (13.5 $\%$) is much smaller ($\sim$100 meV) than that of the prominent peak ($\sim$1 eV).  Therefore, the difference in the energy position of the chemical potential between Co pnictides and LaCo$_{1.73}$Fe$_{0.27}$B$_2$ is mainly due to the difference in the valence electron number between pnictogen and B atoms (three and one $p$ electrons, respectively).  The point (ii) is related to the formation of covalent bonding between the Co 3$d$ and B 2$p$ orbitals as predicted by the previous calculation \cite{Mizoguchi}.  While the As 4$p$ level is much deeper than the transition metal 3$d$ level in pnictides \cite{AsPcalc}, the observed B 2$p$ level is relatively shallow in good agreement with the calculation (see Fig. 1).  The shallow B 2$p$ level prevents the formation of the closed B 2$p$ shell (B$^{5-}$ state) and hybridizes with the Co 3$d$ level to form covalent bonding.  The partially opened shell of B 2$p$ orbitals is supported by rough estimation of the valence: La is +3 and Co is also close to +3 ($\sim$3$d$$^6$) by taking into account the experimental fact that the position of the chemical potential in LaCo$_{1.73}$Fe$_{0.27}$B$_2$ is nearly identical to that in Fe-pnictide superconductors \cite{Singh, Sato}, so that the valence of B is estimated to be -4.5 (2$p$$^{5.5}$).  A wide bandwidth caused by the Co 3$d$-B 2$p$ covalent bonding removes the magnetic moment of Co 3$d$ electrons \cite{Mizoguchi}.  The strong covalent bonding would be also responsible for the observed larger $k_z$ dispersion and the smaller mass renormalization than those of Fe-pnictide compounds.

The next crucial issue is why the partial Fe substitution (hole doping) into the parent LaCo$_2$B$_2$ is necessary to realize superconductivity.  As seen in Fig. 1, the observed asymmetric DOS with respect to $E_{\rm F}$ leads to the enhancement of electronic DOS with lowering of the chemical potential.  Therefore, moderate enhancement of the DOS at $E_{\rm F}$ owing to hole doping, which is favorable for superconductivity, is likely to be a main role of the Fe substitution.  This is consistent with the report that the highest $T_c$ value is achieved in the vicinity of the solubility limit (the highest Fe concentration) \cite{Mizoguchi}.  On the other hand, a small amount of electron doping rather reduces the DOS, in agreement with the absence of superconductivity in electron-doped LaCo$_2$(B$_{1-x}$Si$_x$)$_2$ \cite{Mizoguchi}.  The reported electron-hole asymmetry in the superconducting characteristics may be related to the asymmetric DOS in the vicinity of $E_{\rm F}$.

Finally, we discuss the mechanism of superconductivity.  It is remarked that an electron count in the 3$d$ orbitals of LaCo$_{1.73}$Fe$_{0.27}$B$_2$ is comparable to that of Fe pnictides ($\sim$3$d$$^6$), but different from that of Co pnictides ($\sim$3$d$$^7$).  As described above, we have revealed the multiband nature of superconductivity and the existence of mass-renormalization in LaCo$_{1.73}$Fe$_{0.27}$B$_2$.  These findings suggest that LaCo$_{1.73}$Fe$_{0.27}$B$_2$ is an unconventional superconductor, and a fundamental question arises as to whether a common superconducting mechanism is at work in both Co-boride and Fe-pnictide superconductors.  It has been argued that in Fe pnictides the antiferromagnetic spin fluctuations with $Q$ $\sim$ ($\pi$, 0) enhanced by nesting of the hole and electron FSs at the $\Gamma$(ƒ¤) and M(A) points, respectively, promote a sign-reversal $s_{\pm}$-wave pairing \cite{Mazin, Kuroki}.  This spin fluctuation model has been regarded as a leading candidate for the pairing mechanism in Fe pnictides, while its applicability is still under debate.  As seen in Fig. 4, FS nesting through the $Q$ vector is ill defined in LaCo$_{1.73}$Fe$_{0.27}$B$_2$ due to the size mismatch of the hole and electron FSs at the $\Gamma$(ƒ¤) and M(A) points, respectively.  We do not find other well-defined nesting vectors as well.  The difference between CoB$_4$ and FeAs$_4$ tetrahedral shapes, e.g. the B-Co-B bond angle ($\sim$131$^{\circ}$) \cite{Mizoguchi} is much larger than the As-Fe-As angle in Fe pnictides, and the formation of the aforementioned covalent bonding would cause the different FS topology, even though the number of 3$d$ electrons is comparable between LaCo$_{1.73}$Fe$_{0.27}$B$_2$ and Fe pnictides.  Thus, the present results show that the spin fluctuations are hard to use to explain the superconductivity in LaCo$_{1.73}$Fe$_{0.27}$B$_2$.  Future experimental investigations are required to establish the origin of superconductivity in the LaCo$_2$B$_2$ system.

\section{CONCLUSION}
In conclusion, we reported high-resolution ARPES results on a LaCo$_{1.73}$Fe$_{0.27}$B$_2$ superconductor.  We clarified that the replacement of pnictogen with B atom causes the chemical potential shift and the covalent bonding between the Co 3$d$ and the B 2$p$ orbital.  This modification of the electronic structure moves the system away from the ferromagnetic instability and may realize the superconductivity.  We have also revealed important ingredients relevant to the superconducting mechanism, such as the characteristic shape of the DOS, the moderate electron correlations, and the complex FS topology.  The present results suggest that the spin fluctuation model proposed as a pairing mechanism in Fe-pnictide superconductors is not applicable to the LaCo$_{1.73}$Fe$_{0.27}$B$_2$ superconductor.

\section{ACKNOWLEDGMENTS}
We thank S. Souma, K. Yoshimatsu, H. Kumigashira, and K. Ono for their assistance in ARPES measurements.  We are also grateful to T. Kamiya for his help in the band calculation.  This work was supported by grants from JSPS, TRiP-JST, FIRST, MEXT of Japan, and KEK-PF (Proposal No. 2009S2-005).

\end{document}